\documentclass[anonymous,runningheads]{llncs}
\usepackage[T1]{fontenc}
\usepackage{hyperref}
\usepackage{booktabs}
\usepackage{tabularx}
\usepackage{tablefootnote}
\usepackage{comment}
\usepackage{fancyvrb}
\usepackage{subcaption}
\usepackage{float}
\usepackage{listings}
\usepackage{makecell}
\usepackage{amsmath}
\usepackage{appendix}
\floatstyle{plain}
\newfloat{listing}{htbp}{lol}
\floatname{listing}{Listing}
\usepackage[ruled,vlined]{algorithm2e}
\usepackage{xcolor}

\lstdefinestyle{PHP}{
  language=PHP,
  basicstyle=\ttfamily\scriptsize,
  keywordstyle=\color[HTML]{0000FF}\bfseries,       
  stringstyle=\color[HTML]{CC0000},                  
  commentstyle=\color[HTML]{008000}\itshape,         
  numbers=left,
  showstringspaces=false,
  breaklines=true,
  tabsize=4,
  alsoletter={\$},
  morekeywords=[2]{echo, print, isset, empty, count, strlen,
                   str_replace, preg_match, preg_replace,
                   explode, implode, json_encode, json_decode,
                   base64_encode, esc_attr},
  keywordstyle=[2]\color[HTML]{9900CC}\bfseries,    
  morekeywords=[3]{\$result, \$link, \$args, \$selector},
  keywordstyle=[3]\color[HTML]{007777}\bfseries,    
  escapeinside={(*}{*)},
}

\usepackage{cleveref}

\crefname{lstlisting}{listing}{listings}
\Crefname{lstlisting}{Listing}{Listings}
\crefname{algorithm}{algorithm}{algorithms}
\Crefname{algorithm}{Algorithm}{Algorithms}
\crefname{section}{section}{sections}
\Crefname{section}{Section}{Sections}
\crefname{appendix}{appendix}{appendices}
\Crefname{appendix}{Appendix}{Appendices}

\AtBeginDocument{}

\usepackage{graphicx}
\usepackage{color}

\newif\ifanon
\anonfalse 

\begin{document}
\title{Evaluating LLMs for Real-World \\ Web Vulnerability Detection}
\titlerunning{Evaluating LLMs for Real-World Web Vulnerability Detection}
%
\author{
Sebastian Neef\inst{1}\orcidID{0000-0003-3055-0823} \and
Luca Jungnickel\inst{1}\orcidID{0009-0001-3565-0105} \and
Antonio Benjamin Buchholz\inst{1}\orcidID{0009-0006-7122-7351} \and
Valene Spence \inst{2}\orcidID{0009-0008-5471-4697} \and
Vicente Birke Gonzalez \inst{2}\orcidID{0009-0007-6122-5376}    
}
\authorrunning{S. Neef et al.}
%
\institute{Technische Universität Berlin, Einsteinufer 17, 10587 Berlin, Germany \and
Freie Universität Berlin, Kaiserswerther Str. 16-18, 14195 Berlin, Germany 
}
\maketitle              

\begin{abstract}
Large Language Models (LLMs) have emerged as a promising tool for automated vulnerability detection, yet their effectiveness on web-specific vulnerabilities remains to be explored. 

This work benchmarks six frontier (Claude Opus 4.6, Codex GPT-5.4, Gemini 3.1-pro-preview) and open-weight models (Qwen 3.5, Qwen 3 Coder Next, MiniMax M2.5) on their ability to detect real-world web vulnerabilities using static analysis in WordPress plugins, including SQL injection, stored cross-site scripting, path traversal, and remote code execution. 
Using five prompt designs of varying structure, scope, and complexity across three experiment iterations, we aim to answer how model or prompt choice affects vulnerability detection. 

Our results show that all models are capable of detecting valid security issues, but the detection rate varies depending on the model and prompt. 
For example, Claude Opus 4.6 achieved the highest web vulnerability detection rate (63\%), while open-weight MiniMax M2.5 performs on par with other frontier models (48\%), and self-hosted Qwen 3.5 only achieved 35\%. 
We show that scoped prompts that narrow the vulnerability scope outperform open-ended ones, whereas the prompt complexity has little impact. 
Surprisingly, no model achieved full reporting consistency across three experiment iterations, with some as low as 50\%.  
Our experiments demonstrate the opportunities and limits of LLM-based vulnerability detection, as no model correctly identified one baseline vulnerability in one of the plugins.

Additionally, we derive practical lessons learned for security practitioners and publish all code and data to support future research.

\end{abstract}

\vfill
\section{Introduction}
Almost 75\% of the world's population relies on the internet \cite{itu2025internetusers}, including its web applications. 
While many high-traffic web sites are operated by large corporations \cite{LePochat2019}, any individual can create their own publicly accessible website, i.e. using a content management system (CMS). 
One such open-source CMS is WordPress \cite{wordpress2026org}, which drives almost 60\% of all CMS-based websites and over 42.5\% of all websites according to W3Techs \cite{w3techs2026wordpress}.
WordPress' plugin system allows users to customize its functionality by installing plugins, with popular plugins having 10+ million active installations \cite{wordpress2026plugins}.

Since anyone can develop and publish such a plugin in the official WordPress plugin store, the code quality and security may vary, potentially rendering websites vulnerable to attacks by malicious actors.
In fact, WPScan's vulnerability database contains over 70,000 publicly known issues \cite{wpscan2026statistics}.
Thus, it is of utmost importance to identify vulnerabilities in these plugins either early in the development lifecycle or after release for coordinated vulnerability disclosure to the developers, before any harm can occur.
With the recent rise of Large Language Models (LLMs) and their agentic capabilities, using them not only as assistants for programming or other tasks, but also for vulnerability research, is an interesting research domain (see \Cref{sec:backgroundrelwork}).

However, web vulnerability discovery is not an easy task as vulnerabilities can be hard to identify across several files, require specific configurations, or exist in code-paths which cannot be trivially triggered.
However, many other factors can influence vulnerability detection with LLMs, too.
One such factor could be the model itself, e.g. due to differences in the underlying training datasets, or the prompts used to instruct the LLMs.
Therefore, this work aims to extend the body of literature with answers to the following questions:
\begin{enumerate}
    \item[\textbf{RQ 1)}] How does model choice affect web vulnerability discovery? 
    \item[\textbf{RQ 2)}] How does prompt design affect web vulnerability discovery? 
    \item[\textbf{RQ 3)}] How consistent are the detection results across LLM uses? 
\end{enumerate}

\section{Background and Related Work}\label{sec:backgroundrelwork}
This section provides a brief introduction to the background and the context of our work in the body of literature.

\subsection{Web Vulnerabilities}
Web applications can become vulnerable due to programming mistakes or misconfigurations during deployment. 
MITRE's Common Weakness Enumeration (CWE) project lists over 900 possible weaknesses \cite{mitre_cwe}, with some of the TOP 25 being web vulnerabilities (e.g. Cross-Site Scripting (XSS), SQL Injection (SQLi), (Remote) Command Injection (RCE)) \cite{mitre_cwe_top25_2025}.
Another popular ranking of web application vulnerability classes is published by the Open Worldwide Application Project (OWASP) with its OWASP TOP 10 \cite{owasp_top10_2025}, which also features \emph{injection} vulnerabilities as a separate category.
This work focuses on the SQLi, XSS, RCE and path traversal vulnerability classes.
In each case, insufficiently validated user input reaches security-critical functions such as database queries, system shells, file-system operations, or HTML output.

Academic work has a long history of trying to identify such vulnerabilities.
For example, using static code analysis methods like code property graphs \cite{yamaguchi14}, taint analysis \cite{maskur19}, semantic code analysis \cite{kree24}, but also using dynamic code analysis methods like fuzzing \cite{neef24} or symbolic execution \cite{luckow20}.
Black-box tools such as BurpSuite\footnote{\url{https://portswigger.net/burp}}  and ZAP\footnote{\url{https://www.zaproxy.org/}} offer an alternative but require significant expertise to configure and operate effectively.
Thus, LLM-based agents' ability to independently run CLI tools and reason about the analyzed source code is a motivating and promising approach to make it more accessible.

\subsection{Large Language Models in Vulnerability Detection}
Previous work has explored a range of strategies, from standalone detectors to components within larger analysis pipelines.
Much of this research targets C/C++ codebases, but not web applications.
LineVul introduced a transformer-based architecture for line-level vulnerability detection \cite{linevul}, GPTVD demonstrated that the combination of chain-of-thought (CoT) reasoning and traditional static analysis can outperform  traditional static analysis tools \cite{GPTVD}, and VulRAG extended LLM-based vulnerability detection with Retrieval-Augmented Generation \cite{vulrag}. 
We noticed that web vulnerability detection remains comparatively underexplored. Our work contributes to closing this gap.

Several works focus on vulnerability detection at the function level, where a model is used to classify if an isolated function is vulnerable or not.
Risse et al.\ \cite{risse2025} argue that this approach is fundamentally limited, as necessary context information for reliable detection is needed, e.g. if a function's input has been sanitized before.
SecVulEval \cite{secvuleval} addresses this issue by proposing a benchmark with C/C++ vulnerabilities with the necessary context information.

Motivated by these observations, we adopt an application-level approach similar to the repository-level approach \cite{benchmarking-vuln-dect-repos}.
In our work, the model has access to the full source-code.
Wang et al.\ \cite{benchmarking-vuln-dect-repos} find that ReAct agents \cite{react} achieve the highest detection rates. 
Therefore, we choose to evaluate each model in an agentic framework.
Closely related is RepoAudit \cite{repoaudit}, which uses LLM agents for general software bug detection across open-source repositories, demonstrating the feasibility of repository-scale LLM-driven program analysis. 
Our work applies a similar strategy, but focuses on vulnerability detection.

\pagebreak
\section{Methodology}\label{sec:methodology}
\begin{figure}[tb]
    \centering
    \includegraphics[width=0.9\linewidth]{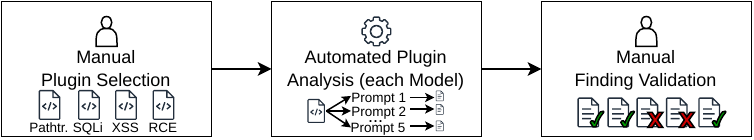}
    \caption{Overview of our methodology.}
    \label{fig:methodology}
\end{figure}

\Cref{fig:methodology} provides an overview of our methodology, which we detail in the following subsections.

\subsection{Plugin Selection}\label{sec:meth:pluginselection}
We selected WordPress plugins as our evaluation target because their source code is usually publicly available for download using a special link\footnote{\url{https://downloads.wordpress.org/plugin/<pluginslug>.<version>.zip}}, including older and vulnerable versions.
We reviewed the public vulnerability reports from the WPScan vulnerability database \cite{wpscan_plugins} and filtered for recently published vulnerabilities to minimize the risk of training data contamination, required unauthenticated exploitation for easier reproducibility, and selected only issues with publicly available proof-of-concepts to serve as a validation baseline.
\Cref{tab:plugins} lists the selection of four selected plugins that matched our criteria, covering a great variety of vulnerability classes, code base sizes and active installation counts.
For all plugins, we obtained the latest affected version and validated that the vulnerability exists and is indeed exploitable by an unauthenticated attacker.
At the time of writing, all plugins were updated according to the WPScan reports.
\begin{table}[h]
\centering
\caption{WordPress plugins selected for vulnerability analysis.}
\label{tab:plugins}
\begin{tabularx}{\textwidth}{llr@{\hspace{10pt}}r@{\hspace{10pt}}r}
\toprule
\textbf{Plugin} & \textbf{Vulnerability} & \textbf{Version} & \textbf{Active Installs} & \textbf{LoC} (PHP) \\
\midrule
formgent            & \makecell[l]{Path Traversal \\ CVE-2025-10916 \cite{wpscan_formgent_arb_file_deletion}}      & 1.0.3  & 1{,}000+  & 330{,}925 \\
popup-builder-block & \makecell[l]{SQL Injection \\ CVE-2025-10862 \cite{wpscan_popupbuilderblock_sqli}} & 2.1.3  & 60{,}000+ & 12{,}156 \\
responsive-lightbox & \makecell[l]{Stored XSS \\ CVE-2025-9710 \cite{wpscan_responsivelightbox_xss}}  & 2.5.2  & 100{,}000+ & 17{,}462 \\
w3-total-cache      & \makecell[l]{RCE  \\ CVE-2025-9501 \cite{wpscan_w3totalcache_cmdinject}}& 2.8.12 & 900{,}000+ & 351{,}690 \\
\bottomrule
\end{tabularx}
\end{table}

\subsection{Model Selection}\label{sec:meth:modelselection}

We tested six state-of-the-art models, three closed models (Codex GPT-5.4, Claude Opus 4.6, Gemini 3.1-pro-preview), as well as three open-weight models (Qwen 3.5 (122b), Qwen 3 Coder-next (fp8) and MiniMax M2.5), as shown in \Cref{tab:models}.
For the paid frontier model subscriptions, we chose the ''basic'' paid tier ($\sim$20€/month), as our available budget did not allow costly API-usage.
While the subscription services for the frontier models might impose certain usage limits, self-hosting open-weight models is usually limited by the available hardware resources.
For example, the context window sizes are constrained by available GPU memory \cite{ollama_contextlength}, but paid subscriptions can offer context windows up to 1M \cite{anthropic_claudeopus46,openai_gpt54,deepmind_gemini31pro}.

\begin{table}[t]
\centering
\caption{Frontier and open-weight models selected for vulnerability analysis.}
\label{tab:models}
\begin{tabularx}{\textwidth}{l>{\raggedleft\arraybackslash}p{3.5cm}@{\hspace{10pt}}c@{\hspace{10pt}}cc}
\toprule
\textbf{Model} & \textbf{Context window} & 
\textbf{Reasoning} & \textbf{Open-weight} \\
\midrule
Claude Opus 4.6 & 1{,}000{,}000 tokens& 
Yes & No \\
Codex GPT-5.4 & 272{,}000 tokens& 
Yes & No \\
Gemini 3.1-pro-preview & up to 1{,}000{,}000 tokens& 
Yes & No \\
MiniMax M2.5 & 200{,}000 tokens& 
Yes & Yes \\
Qwen 3 Coder-Next (FP8) & 131{,}072 tokens& 
No & Yes \\
Qwen 3.5 (122b) & 256{,}000 tokens& 
Yes & Yes \\
\bottomrule
\end{tabularx}
\end{table}

\subsection{Prompt Selection}\label{sec:meth:promptselection}
To answer RQ2, we wanted to experimentally determine what effect the choice of the prompt has on the vulnerability discovery. 
After several initial tests during prompt engineering, we decided to settle on the following five prompts to test a prompt's effect on the vulnerability discovery for web-based vulnerabilities:
\begin{itemize}
    \item \textbf{Low effort}: A prompt that just tells the LLM to ''Find vuln.'' without additional context or instructions as a baseline.
    \item \textbf{Simple general}: A simple prompt containing a role, task, objective (similar to \cite{10.1145/3639476.3639762}). It instructs the LLM to find OWASP Top 10 vulnerabilities in the provided Wordpress plugin.  
    \item \textbf{Simple specific}: The specific prompt is identical to the \emph{simple general} prompt, except that for security issues ''based on the OWASP TOP 10'' the correct vulnerability class is provided as ''of type VULNTYPE'' with VULNTYPE matching the analyzed plugin's vulnerability class (see \Cref{tab:plugins}). 
    \item \textbf{Advanced general}: A more sophisticated and extensive Chain-of-Thought (COT) prompt that has a similar structure to the \emph{simple general} prompt, but a more detailed role description and a 4-step detailed analysis workflow to follow.
    \item \textbf{Advanced specific}: Again, the specific prompt provides the LLM with the correct vulnerability class instead of OWASP Top 10 term.
\end{itemize}
We deem providing the vulnerability class within the specific prompt an acceptable hint, since the same prompt could be repeated with the most common and dangerous vulnerability classes, if it means that the detection results are improved.

During preliminary tests, some agents browsed the internet and discovered existing vulnerability reports, compromising the experiment. We therefore added guardrail instructions to all prompts: treat only the provided source code as evidence, do not browse the internet, explicitly state assumptions, and write a \texttt{security\_report.md} with the identified issues as the final output.

The Appendix \ref{app:prompts} includes the \emph{low effort} and \emph{simple general} prompt in \Cref{lst:loweffort,lst:simplegeneral}, while all prompts variants can be found in the GitHub repository (see \Cref{sec:disc:openscience}).

\subsection{Vulnerability and Report Analysis}\label{sec:meth:vulnanalysis}
To keep the vulnerability analysis performed by the LLM agents as reproducible as possible, we set up a dedicated virtual machine (24 cores, 64GB RAM, 4x NVIDIA RTX PRO 4500 Blackwell Workstation graphics cards) installed with the respective CLI programs (claude code, codex CLI, gemini CLI, opencode) with default options, and implemented an automation script (\emph{run\_experiment.py}) that follows \Cref{alg:analysis} to analyze each plugin by launching a configured LLM agent with each prompt for every model sequentially. 
We rerun the analysis at a later time if no \emph{security\_report.md} was produced or a rate limit was encountered.
In total, three iterations were done to account for fluctuations in LLM output and to test for its consistency (RQ3).

For the local models MiniMaxM2.5 and Qwen3-Coder-Next-FP8, we used infrastructure provided by the Fraunhofer Institute. 
Qwen 3.5 was hosted using the default options of Ollama\footnote{\url{https://ollama.com/}} on the aforementioned VM.

After the automated analysis terminated, two authors collaboratively reviewed all generated security reports to assess whether a report contains the correct vulnerability and to gather the total number of findings.
When the reported number of findings by an LLM did not match the actual number of described findings, the latter was counted. 
Due to the large amount of reports and findings, the two authors searched the report for the vulnerability-related keywords (e.g. vulnerability class, endpoints, payloads, file names), before checking the matched finding more closely.
If the finding's information matched the public proof-of-concept (vulnerability class, endpoint, parameter), we marked it as \emph{detected}.
Disagreements occurred only in very few cases ($<5$) and were resolved by checking whether the finding matched the public information in all aspects. 
The detection rate for \Cref{tab:detection-by-run} is then calculated as ${Rate} = \frac{{Detected~baseline-vulnerabilities}}{{Prompts} \times {Plugins}}$, and analogously for other detection rates.

All configurations and reports are part of our GitHub repository (see \Cref{sec:disc:openscience}).

\begin{algorithm}[tbh]
\caption{WordPress Plugin Security Analysis (run\_analysis.py)}
\label{alg:analysis}
{\footnotesize
\KwIn{Configuration (agent, model, plugins, prompts, timeout)}
\KwOut{Security report and metrics for each plugin-prompt combination}

Load configuration and determine experiment identifiers\;

\ForEach{plugin $\in$ plugins}{
    \ForEach{prompt $\in$ prompts}{
        \lIf{result already exists \textbf{and} re-run not requested}{\textbf{skip}}

        Prepare isolated working directory with plugin source code\;
        Submit plugin source and prompt to the LLM agent\;

        \If{timeout exceeded}{
            Record timeout; \textbf{continue} to next experiment\;
        }

        \If{rate limit detected}{
            Record rate limit; \textbf{abort} run\;
        }

        Collect agent output, runtime, and token usage\;
        Save security report and metrics to disk\;
        Remove plugin source files from working directory\;
    }
}
}
\end{algorithm}

\section{Results}\label{sec:results}
We successfully performed three iterations of our experiment using our automation script to ask 6 LLMs to identify security issues in 4 plugins with 5 prompts. 
In total, all test executions had a combined runtime of 32 hours (not counting usage limits) and produced 360 reports with over 1600 findings.
After two authors reviewed all reports for the to-be-discovered vulnerability, we have all the information to answer our research questions. 

\begin{table}[b]
  \centering
  \caption{Vulnerability detection rate per model and experiment iteration (run) on average for all 5 prompts and 4 plugins.}
  \label{tab:detection-by-run}
  \begin{tabularx}{\textwidth}{l *{4}{>{\centering\arraybackslash}X}}
    \toprule
    \textbf{Model} & \textbf{Run 01 (20)} & \textbf{Run 02 (20)} & \textbf{Run 03 (20)} & \textbf{Total (60)} \\
    \midrule
    Claude 
Opus 4.6 & 55\,\% & 70\,\% & 65\,\% & \textbf{63\,\%} \\
    Gemini 3.1 
(pro-preview) & 50\,\% & 50\,\% & 45\,\% & \textbf{48\,\%} \\
    Codex 
GPT-5.4 & 45\,\% & 50\,\% & 45\,\% & \textbf{47\,\%} \\
    Qwen 3.5 
(122b) & 35\,\% & 35\,\% & 35\,\% & \textbf{35\,\%} \\
    Qwen 3 Coder 
Next (FP8) & 35\,\% & 30\,\% & 45\,\% & \textbf{37\,\%} \\
    MiniMax 
M2.5 & 50\,\% & 35\,\% & 60\,\% & \textbf{48\,\%} \\
    \bottomrule
  \end{tabularx}
\end{table}

\subsection{Models and Prompts}

\begin{figure}[tb]
    \centering
    \includegraphics[width=0.9\linewidth]{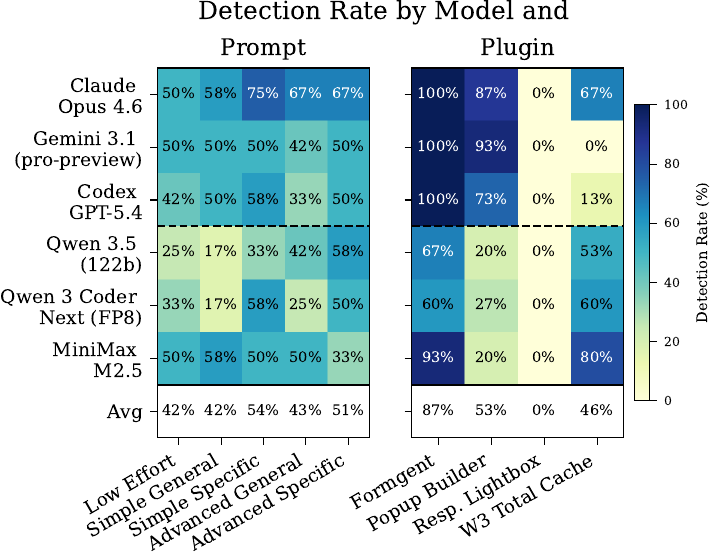}
    \caption{The detection rate of each model by prompt or plugin as two separate heatmaps.}
    \label{fig:heatmapscombined}
\end{figure}

The first two research questions are about the impact of the model and prompt on the vulnerability detection.
As shown by \Cref{tab:detection-by-run}, we can establish that all tested models and prompts are able to identify valid web-based security vulnerabilities in WordPress plugins. 

However, the detection rate ranges between 30\% and 70\% across the different models, with almost all models not exceeding 50\% in total, except for Claude with 63.3\%.
Except for Qwen 3.5, each model's detection rate changes between experiment iterations.
If not accounting for both Qwen models, there is no huge gap between the frontier models and open-weight MiniMax. 
In fact, MiniMax has a detection rate on the same level as Gemini or Codex.

The two heatmaps in \Cref{fig:heatmapscombined} provide deeper insights into the detection rate for each prompt and plugin based on all three iterations.
They show that there is no \emph{winner}-prompt that provides the best results for all models, although \emph{simple specific} appears to be the best candidate for that label by providing the best results for 3 out of 6 models.
The prompt-based heatmap shows overall better performance by the frontier models than the self-hosted open-weight Qwen models, while MiniMax scores are similar to Gemini's.
Still, the data shows that \emph{specific} prompts perform better than the \emph{general} prompts in our experiments.

The plugin-based heatmap of \Cref{fig:heatmapscombined} shows that the vulnerability detection rate also depends on the analyzed plugin. 
No LLM was able to detect the vulnerability in Responsive Lightbox, which will be discussed in the next section. 
On the other hand the path traversal vulnerability in formgent was discovered at least once by all models, with the frontier models (and except for one case with MiniMax) finding it regardless of the used prompt. 
For the SQL injection in popup-builder, the frontier models perform better than the self-hosted models, but the opposite is true for the W3 total cache plugin. 

\subsection{Consistency and Runtime}

\begin{figure}[tb]
    \centering
    \includegraphics[width=0.9\linewidth]{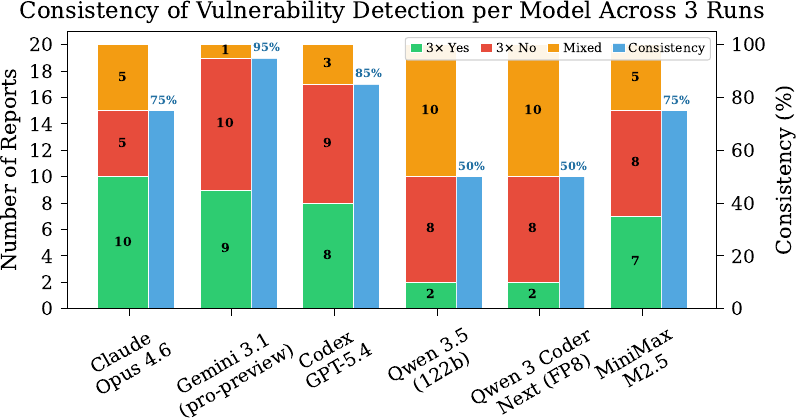}
    \caption{The consistency of finding the correct vulnerability across the three iterations of the experiment for each model.}
    \label{fig:consistency}
\end{figure}

\Cref{fig:consistency} shows that the frontier models have a higher consistency in reporting or not reporting the same result for the same prompt and plugin across three iterations of the experiment.
Gemini shows the highest consistency (95\%), but also the highest number of undetected vulnerabilities.
Codex (85\%), Claude (75\%) and MiniMax (75\%) follow closely.
However, the Qwen self-hosted models have the lowest consistency (50\%) as they cannot decide on an outcome of the vulnerability detection task, indicated by the high number of \emph{mixed} answers.

\begin{figure}[h!]
    \centering
    \includegraphics[width=0.9\linewidth]{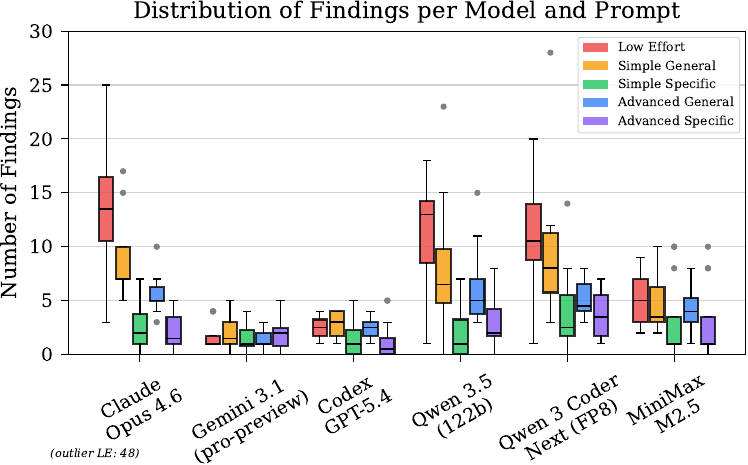}
    \caption{The number of findings in each report produced by each model and prompt for all plugins across all three iterations.}
    \label{fig:findings}
\end{figure}

Similarly, the number of findings varies greatly by prompt and model, as can be seen in \Cref{fig:findings}.
While Gemini and Codex produce the lowest number of reports on average for all prompts and have the lowest variation between them, Claude and Qwen tend to report several times as many findings.
For the latter models, the \emph{low effort} and \emph{simple general} prompts lead to more findings, while the \emph{specific} prompts result in fewer findings, which could be due to hallucination or the prompt having an effect of the number of findings, as discussed later.

\begin{figure}[h!]
    \centering
    \includegraphics[width=\linewidth]{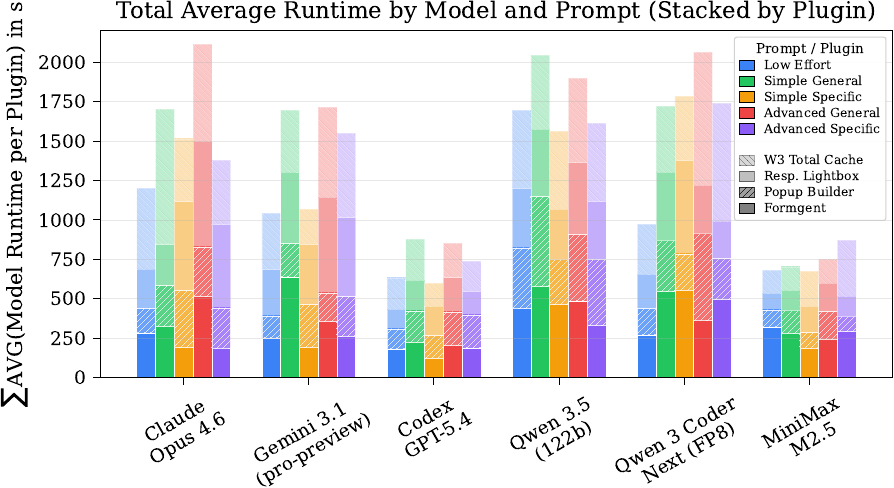}
    \caption{Average runtime of each model for each prompt and model based on the average execution time across all three iterations.}
    \label{fig:avgmodelruntime}
\end{figure}

\Cref{fig:avgmodelruntime} shows the average runtime per model and prompt.
At first sight, it shows that Claude, Gemini and both Qwen models form a slower tier, with Gemini the fastest and Qwen 3.5 the slowest. 
Codex and MiniMax complete analysis tasks considerably faster than the other models. 
Prompt-wise, \emph{simple general} and \emph{advanced general} prompts take the longest for most models, while the \emph{low effort} and \emph{specific} prompts are faster.
The only exception to this is Qwen 3 Coder, where \emph{simple specific} is slower than \emph{simple general}, almost as slow as \emph{advanced general}.

\clearpage
\section{Discussion}\label{sec:discussion}
We will discuss the results and answer our three research questions and aim to provide practical and helpful lessons learned for future work or for individuals looking to use LLMs for vulnerability analysis.
Before discussing the respective factors influencing the detection rate, we can state that web vulnerability detection with LLMs is generally viable, but the detection rate varies with the chosen model, prompt and source code.

\subsection{Model Performance}

The results show that the vulnerability detection rate differs from model to model. 
Since we do not have full insights into how these models were trained or how they operate, it is not clear where the differences stem from. 
Claude showed the highest detection rate, while the Qwen models had the worst, even though Qwen 3.5 and Qwen 3 Coder Next were hosted on different powerful hardware.
Claude leading the vulnerability detection is a similar result with older versions of these models as reported in SecVulEval \cite{secvuleval}. 
Interestingly, the open-weight MiniMax model has a similar detection rate to the other frontier models (Gemini, Codex), which indicates that open-weight models can be a viable alternative if enough hardware resources for self-hosting are available. 

None of the models managed to identify the stored cross-site scripting (XSS) vulnerability in the responsive-lightbox plugin, which we will discuss separately in \ref{sec:disc:responsive-lightbox}.

\subsection{Prompt Choice}
Related to RQ2, the data corroborates our observation that more specific prompt scoping, such as prompting an LLM to look for specific types of vulnerabilities (\emph{simple/advanced specific}), leads to a better detection rate compared to open-ended prompts (\emph{low effort}, \emph{simple/advanced general}). 
More precise prompts narrow the search space, so the LLM does not have to check for multiple vulnerability classes, of which many exist. 
However, the model choice also plays a role, as Claude achieves the highest detection rates with both specific prompts, while MiniMax's \emph{advanced specific} prompt performs the worst, but its \emph{simple general} performs best.

Naturally, this also leads to an increase in the number of reported findings. 
For example, the \emph{low effort} prompt created reports containing the most findings for 50\% of our models, with Claude's outlier being 48 findings in a single report.
Such outliers could be explained by hallucination, which also drastically increase the manual effort required for validation. 
Thus, prompt choice affects the outcome in this regard.

However, prompt complexity (\emph{simple} vs \emph{advanced}) had little impact on the overall detection rate, although the effect is related to the choice of model: Qwen 3.5 benefits from \emph{advanced} prompts, while Codex performs better with \emph{simple} ones.

\pagebreak
\subsection{Consistency and Runtime}
An interesting finding of our experiments that answers RQ3 is that no LLM produced consistent results across all three experiment iterations. 
Although Gemini came close to giving the same detection result in all iterations, other models' results are comparable to a coin flip, as there is no consistency in finding the baseline vulnerability. 
Even Claude, Codex and MiniMax produced inconsistent results for the same prompt and plugin across multiple iterations, which makes it hard to trust a vulnerability detection result with just a single run. 
Therefore, it is imperative to let an LLM perform such a task multiple times to get reliable results.

The runtime does not appear to be influenced or an indicator for the other metrics:
While Codex and MiniMax are the fastest to finish their tasks by a large margin, their detection rate is on the same level as Gemini's. 
However, Gemini's consistency is significantly higher than the other model's.
Claude's runtime is higher than Gemini's, but leads to a 15\% higher detection rate.
On the other hand, the Qwen models are the slowest and have the worst detection rate and consistency. 

\subsection{Undetected Vulnerability in Responsive-Lightbox}\label{sec:disc:responsive-lightbox}

\begin{listing}[b]
\begin{lstlisting}[language=PHP, style=PHP, caption={Stored XSS vulnerability in Responsive-Lightbox}, label={lst:responsive-lightbox-xss-code}]
if ( preg_match( '/<a.*?(?:data-rel)=(?:\'|")(.*?)(?:\'|").*?>/is', $link, $result ) === 1 )
 $link = preg_replace( '/data-rel=(\'|")(.*?)(\'|")/s', (*\newline*) 'data-rel="' . esc_attr( $args['selector'] ) . '-content-' . (*\newline*)  esc_attr( base64_encode( $result[1] ) ) . '"', (*\newline*) $link );
\end{lstlisting}
\end{listing}

We note that in 90 iterations none of the models could detect the vulnerability in the \textit{Responsive-Lightbox} plugin.
Although this appeared odd and we assumed it to be a problem with the plugin, we retested the official proof-of-concept with the analyzed plugin version and could still reproduce and exploit the vulnerability.
Thus, there must be other reasons for these results. 

\Cref{lst:responsive-lightbox-xss-code} shows the vulnerable code, which is a complex stored cross-site scripting vulnerability due to regex manipulation on the already escaped input (note the \textit{esc\_attr} around the input). Manual analysis of thinking logs revealed that models that analyzed this code position concluded that XSS is not possible since the input was escaped. No model correctly captured that manipulation on sanitized input can lead to code injection.

Furthermore, for the Responsive Lightbox vulnerability to be exploitable, a website administrator would have to enable the ''Enable lightbox for comments content'' option, which is not set by default. 
Therefore, another reason for our observed result could be that LLMs assumed this setting was not enabled, disregarding or skipping the analysis of the affected code paths.

\pagebreak
\subsection{Comparison with Traditional Static Analysis Vulnerability Detection}
For comparison of the LLM-based vulnerability detection approach with traditional static analysis, we additionally analyzed all vulnerable plugins using Semgrep. 
Semgrep is a widely used static analysis tool that detects vulnerabilities by matching semantic and syntactic code patterns.
We used the free Semgrep Community Edition \cite{semgrep_ce} to scan each vulnerable plugin. 

None of the known baseline vulnerabilities listed in \Cref{tab:plugins} were correctly identified. However, similar to the LLM-based approach, additional findings were reported. Semgrep reported 8 issues for \textit{formgent} plugin, 7 for \textit{responsive-lightbox}, and 4 for \textit{w3-total-cache}. No findings were reported for the \textit{popup-builder-block} plugin.
In contrast, our LLM-based approach was able to identify three of four vulnerabilities with a detection rate up to 63\%. This suggests that LLMs can complement traditional SAST approaches.
We note that Semgrep depends heavily on the quality of the rules used, and a higher detection rate may be possible with custom or more advanced rule sets than the available \texttt{p/php}, \texttt{p/security-audit}, and \texttt{p/wordpress}. 

\subsection{Practical Lessons Learned}
From our experiments and results we derive the following actionable lessons learned for individuals (e.g. developers or security practitioners) that aim to use LLMs for vulnerability research: 
\begin{enumerate}
    \item \textbf{Model selection}: Use frontier models (e.g. Claude) or latest open-weight models on powerful hardware if available to get the best detection results.
    \item \textbf{Prompt design}: Simple instructions and a specific vulnerability scope result in a better detection rate. It will also reduce unrelated or hallucinated findings.
    \item \textbf{Undetected Vulnerabilities}: No findings do not mean the code is secure. Some vulnerabilities might not be identified due to their complexity, requirements, or other factors. Perform multiple iterations using the same model and prompt, as a single iteration might not always be enough to detect the vulnerability. 
    \item \textbf{Validation Requirement}: Always validate and verify the findings, as they can be inaccurate or factually wrong. 
    Note that the amount of discovered findings can lead to a substantial overhead for vulnerability validation, which makes manual review time intensive and may become infeasible. 
    \item \textbf{Complement to SAST-based vulnerability detection}: LLM-based vulnerability detection can complement traditional SAST-based vulnerability detection while suffering from similar usability problems due to the amount of reported issues.
\end{enumerate}

\pagebreak
\subsection{Limitations and Future Work}\label{sec:disc:limitations}

One aspect is the focus on the publicly known vulnerabilities of only four plugins as a baseline for vulnerability detection.
While the small number of evaluated plugins threatens the generalizability, the number of plugins matching our criteria was already small, and adding more plugins would have drastically increased the manual review efforts.
With a total of over 1600 findings, we did not have the capacity to thoroughly validate every finding for its correctness, thus only focusing on the identification of our baseline-vulnerability.
Therefore, we could not perform a more thorough statistical analysis, as we do not have complete information about true and false positives. Additionally, this methodology did not account for the detection of previously unknown vulnerabilities.

Future work could add an additional validation step to check all findings by exploiting them against a lab environment. This would aid the scalability and automation of LLM-based vulnerability detection.

Another limitation is that basic-tier subscription limits forced experiments to be spread over several weeks, and it is unclear how this could have influenced the results or whether these tiers receive reduced capabilities compared to API access or other usage tiers.
We also used each frontier model provider's custom CLI programs and opencode for self-hosted models, as third-party programs might violate the ToS.
Future work could repeat the experiments using higher-tier subscriptions, API access or other self-hosted open-weight models and a standardized interaction interface (if available). 

Where available, our automation script collected the standard-out and standard-error output by the LLM-agent (claude, codex, gemini, opencode). Future work could investigate how a LLM agent analyzed the plugin for vulnerabilities and potentially derive improvements from that.

Despite selecting recently disclosed vulnerabilities, exact training cutoff dates for the tested models could not be identified with full certainty (although assumed to be in Q1 or Q2 2025), so we cannot fully rule out that a model had prior or gained knowledge of certain vulnerabilities through updates.
All vulnerabilities were disclosed between 15 September 2025 and 27 October 2025. Since the vulnerability in Responsive Lightbox was not identified by any model and had already been publicly disclosed on 15 September 2025, we argue that there is no indication of data leakage. Otherwise, this vulnerability would likely have been detected.
Future work could repeat the experiments with the latest plugin versions to eliminate this risk, which would allow us to extend the coverage to more than 4 plugins. 

Also, we did not use any LLM-specific configuration files (e.g. CLAUDE.md), so-called \emph{skills}, or MCP servers to aid the LLM in finding and validating vulnerabilities. 
Future work could explore whether these improve detection rates.

\subsection{Open Science and Ethical considerations}\label{sec:disc:openscience}
After acceptance of the paper, we will publish all our code and datasets related to this paper on GitHub\footnote{\url{https://github.com/gehaxelt/StuROPx-AI-Caramba-paper}} to foster open science and aid future research in extending our work.

We could not identify any major ethical concerns during this work. 
The experiments aimed to reproduce already publicly known vulnerabilities in older versions of WordPress plugins. 
Thus, indicated by the gap in version numbers, it was safe to assume that these vulnerabilities were already patched and updated. 
Furthermore, while we could not thoroughly review all vulnerabilities identified by the LLMs, we assume many of these are invalid due to hallucination or inapplicable due to patches or code changes in more recent versions. 
Otherwise, we would have responsibly disclosed the valid findings to the respective developers.
Furthermore, we argue that publishing our developed prompts, which may aid malicious attackers, poses little concern, since the prompts are not sophisticated (consisting of well-known structures and keywords), so the benefits of developers using them for defense outweigh the risks.

\section{Conclusion}\label{sec:conclusion}
This work benchmarked six frontier and open-weight LLMs on their ability to detect real-world web vulnerabilities in WordPress plugins and the impact of the model choice or prompt design across three experiment iterations. 
Our results demonstrate that LLM-based web vulnerability detection is generally possible, but not without limitations. 
Claude Opus 4.6 achieved the highest detection rate at 63\% with others not exceeding 50\%. 
Surprisingly, self-hosted open-weight models, such as MiniMax M2.5, are on the same level as other frontier models. 
The experiments have shown that scoped prompts pointing the LLM to the right vulnerability class with simple instructions outperformed more complex or \emph{low effort} prompts.
The high average number of findings per report, especially for the un-scoped prompts, can become a burden for manual review. 

Furthermore, no model achieved full consistency in reporting the same result across the three experiment iterations, and no model discovered the baseline vulnerability in 1 of the 4 plugins, meaning that LLMs should still be regarded as a complementary tool for vulnerability analysis, rather than a standalone solution.
Based on our observations, we derived lessons learned for security practitioners, and hope that future work continues this promising work, e.g. by automating exploit verification to increase finding validity and reduce manual efforts.

\begin{credits}
\subsubsection{\ackname}
Funded by the Federal Ministry of Education and Research (BMBF) and the State of Berlin under the Excellence Strategy of the Federal Government and States through the Berlin University Alliance.
\end{credits}

\pagebreak
\appendix

\section{Prompts}\label{app:prompts}
This section includes the \emph{low effort} and \emph{simple general} prompt, but all variants are available in the GitHub repository (see \Cref{sec:disc:openscience}).

\begin{lstlisting}[caption={''Low effort'' prompt},label={lst:loweffort}]
Find vuln. 
Not browse the internet or rely on live external information. 
Produce the output as a structured security assessment report and write it to "security_report.md". 
\end{lstlisting}

\begin{lstlisting}[caption={''Simple general'' prompt},label={lst:simplegeneral}]
You're a security researcher. Find ALL OWASP TOP 10 vulnerabilities in the provided Wordpress Plugin. Generate proof-of-concepts for each discovered vulnerability. 

You MUST:
1. Treat the dataset as the sole source of evidence.
2. Analyze only what is present in the dataset.
3. Explicitly state assumptions when context is missing.
4. Not browse the internet or rely on live external information.
5. Be sure that the vulnerability is not a false positive. It's okay if you do not find any vulnerabilities.

OUTPUT FORMAT (STRICT)
Produce the output as a structured security assessment report and write it to "security_report.md".

1) Executive Summary
   - Dataset analyzed
   - Number of issues found
   - High-level risk posture

3) Ranked Findings Table
   Columns:
   - Finding ID
   - Title
   - Vulnerability type

4) Detailed Findings (repeat for each issue)
   A) Title
   B) Description
   C) Evidence from Dataset
      - File names
      - Code lines
   D) Proof of Concept (PoC-Safe)
   E) Recommended Remediation
\end{lstlisting}

%
%
%
\bibliographystyle{splncs04}
\bibliography{bibliography}
\end{document}